\documentclass[10pt]{article}
\usepackage{multicol}
\usepackage{graphicx}
\usepackage{amsmath}
\usepackage[a4paper]{geometry}
\usepackage{hyperref}
\usepackage{rotating}

\begin{document}

\begin{center}
{\Large \bf Analyses of the collective properties of hadronic matter in Au-Au collisions at 54.4 GeV}

\vskip1.0cm

M.~Waqas$^{1}${\footnote{Email (M.Waqas):
waqas\_phy313@yahoo.com; waqas\_phy313@ucas.ac.cn}}, G. X. Peng$^{1,2}${\footnote{Corresponding author. Email (G. X. Peng): gxpeng@ucas.ac.cn}},
M. Ajaz$^{3}${\footnote{Corresponding author. E-mail: ajaz@awkum.edu.pk; Muhammad.Ajaz@cern.ch}}, A. Haj Ismail$^{4, 5}$, %{\footnote{E-mail: a.hajismail@ajman.ac.ae}},
E. A. Dawi$^{4, 5}$%{\footnote{E-mail: e.dawi@ajman.ac.ae}} 
\\

{\small\it  $^1$ School of Nuclear Science and Technology, University of Chinese Academy of Sciences,
Beijing 100049, China

$^2$ Theoretical Physics Center for Science Facilities, Institute of High Energy Physics, Beijing 100049, China

$^3$ Department of Physics, Abdul Wali Khan University Mardan, Mardan 23200, Pakistan

$^4$ Department of Mathematics and Science, Ajman University, Ajman 346, United Arab Emirates

$^5$ Nonlinear Dynamics Research Center (NDRC), Ajman University, Ajman 346, United Arab Emirates
}

\end{center}

\vskip1.0cm

{\bf Abstract:} We investigated the strange hadrons transverse momentum ($p_T$) spectra in Au-Au collision at $\sqrt {s_{NN}}$ = 54.4 GeV in the framework of modified Hagedorn function with embedded flow. It is found that the model can describe the particle spectra well. We extracted the kinetic freeze-out temperature $T_0$, transverse flow velocity $\beta_T$, kinetic freeze-out volume $V$, mean transverse momentum $<p_T>$, the entropy parameter $n$ and the multiplicity parameter $N_0$. We reported that all these parameters increase towards the central collisions. The larger kinetic freeze-out temperature , transverse flow velocity, kinetic freeze-out volume and the entropy parameter (n) in central collisions compared to peripheral collisions show the early decoupling of the particles in central collisions. In addition, all the above parameters are mass dependent. The kinetic freeze-out temperature ($T_0$), the entropy parameter $n$ and mean transverse momentum ($<p_T>$) are larger for massive particles, while the transverse flow velocity ($\beta_T$), kinetic freeze-out volume ($V$) and the multiplicity parameter ($N_0$) show the opposite behavior. Larger $T_0$, $n$ and smaller $\beta_T$ as well as $V$ of the heavier particles indicates the early freeze-out of the heavier particles, while larger $<p_T>$ for the heavier particles evince that the effect of radial flow is stronger in heavier particles. The separate set of parameters for each particle shows the multiple kinetic freeze-out scenario, where the mass dependent kinetic freeze-out volume shows the volume differential freeze-out scenario. We also checked the correlation among different parameters, which include the correlation of $T_0$ and $\beta_T$, $T_0$ and $V$, $\beta_T$ and $V$, $<p_T>$ and $T_0$,  $<p_T>$ and $\beta_T$,  $<p_T>$ and $V$, $n$ and $T_0$, $n$ and $\beta_T$, and $n$ and $V$, and they all are observed to have positive correlations with each other which validates our results.
\\



\vskip1.0cm

\begin{multicols}{2}

{\section{Introduction}}
Heavy ion collisions at the ultra-relativistic energies provide a strong evidence [1--5] for a novel phase of Quantum chromodynamics (QCD) matter, which is called Quark gluon plasma (QGP) [6--10]. Many approaches [11--23] have been developed to study the QGP matter in order to understand the laws of nature at sub-atomic level, but due to its short lifetime its direct measurement is very difficult. However, the dynamical models can play a vital role to explore this provoking and exciting phenomenon happening in the early stages of interaction. Nonetheless, it is understandable that the indirect investigation depends on the excellence of the model to repeatedly and precisely re-establish the related dynamics and phases of the collision from hadron formation to their detection.

Temperature (T) versus baryon chemical potential ($\mu_B$) is usually plotted to show the phase diagram of QCD. In heavy ion collisions, the evaluation of particle yields and their resemblance to the statistical models suggest that $T$ and $\mu_B$ differ in an inverse manner with the center of mass energy ($\sqrt {s_{NN}}$) at the chemical freeze-out [24]. $T$ increases with $\sqrt {s_{NN}}$ while $\mu_B$ decreases [25]. Therefore by varying $\sqrt {s_{NN}}$, the two axes of the phase diagram $T$ and $\mu_B$, can be changed, and get access to a large part of the phase space experimentally. For the study of the phase structure of QCD phase diagram, the Beam Energy Scan (BES) program [26--28] has been designed. Among the different $\sqrt {s_{NN}}$, the STAR experimental group also collected high statistics data for Au-Au collisions at 54.4 GeV in 2017, which allows to extend the accurate measurements of different observables especially from intermediate to high $p_T$ [29].

There are many kinds of temperatures studied in literature which includes the initial temperature ($T_i$) [30, 31], chemical freeze-out temperature ($T_{ch}$) [32--34], effective temperature ($T$) [35--38] and kinetic freeze-out temperature ($T_0$) [39--42]. These temperatures correspond to different stages of evolution. Since the selection of the modified Hagedorn function with embedded flow [43] in the present work due to its closeness to the ideal gas model is brought under consideration, which describes the final state properties of the particles. Therefore we are not interested in other temperatures except the kinetic freeze-out temperature.

Kinetic freeze-out of particles is a very complex phenomenon. Different literatures provide the evidence of different kinetic freeze-out scenarios. For instance ref. [44] shows the single kinetic freeze-out scenario, ref. [45--47] shows double kinetic freeze-out scenario, and similarly ref. [48--50] shows the multiple kinetic freeze-out scenario. In our recent work [51], we also observed a triple kinetic freeze-out scenario, where we studied non-strange, strange and multi-strange particles and found that the multi-strange particles freeze-out early than the non-strange and strange particles. It also increases our curiousity to study the charm particles in future, and we expect their earlier freeze-out than the multi-strange particles. Though there are different opinions about the freeze-out scenarios but it is still an open question in the community. Furthermore, the dependence of the kinetic freeze-out temperature ($T_0$) from central to peripheral collision has different studies. Several literatures claim the decrease of $T_0$ with increasing centrality [52, 53] which suggests the longer lived fireball in central collisions. However, according to some literatures [31, 39, 50, 21, 54, 55], $T_0$ decrease with decreasing centrality which suggests the higher degree of excitation of the system in central collisions. This is also an open question in the community up to now. Both the trends of increasing or decreasing of $T_0$ with centrality are correct in their own explanations.

The transverse momentum ($p_T$) spectra of the particles are very significant observables in high energy collisions, and they can be used for the scrutiny of the dynamics of the particles production. In the present work, we analyze the $p_T$ spectra of strange particles ($k_s^{0}$, $\Lambda$, $\bar {\Lambda}$, $\bar{\Xi}^{+}$ and $\Xi^{-}$) by the modified Hagedorn function with embedded flow [43], and extracted the bulk properties of matter in terms of kinetic freeze-out temperature, transverse flow velocity and kinetic freeze-out volume.

The remainder of the paper consists of method and formalism followed by the results and discussions and then conclusions.
\\

{\section{The method and formalism}}
In high energy collisions, it is a known fact that the high $p_T$ part of transverse momentum spectra of the particles is well described by the quantum chromodynamics (QCD) inspired Hagedorn function [56]

\begin{equation}
\label{eq:1}
\frac{1}{N}\frac{d^2N}{2\pi p_Tdp_Tdy} = C(1+\frac{m_T}{p_0})^{-n}
\end{equation}
 Where {\bf $N$ represents the number of charged particles of a particular type, $p_T$ is the transverse momentum spectra}, "C" is constant of normalization, $p_0$ and $n$ are the free parameters, $m_T$ is the transverse mass and can be expressed as $m_T = \sqrt{(m_0)^2 +(p_T)^2}$ and $m_0$ is the rest mass of the hadron specie.

 As we know that the invariant $p_T$ and $m_T$ distribution of the particles in high energy collisions at RHIC and LHC [57--64] can be well described by the Tsallis distribution function [65, 66]. There are many versions of Tsallis function [64, 67--71] and its more simplified version at mid-rapidity is

 \begin{equation}
\label{eq:1}
\frac{1}{N} \frac{d^2N}{2\pi p_Tdp_Tdy} = C(1+(q-1)\frac{m_T}{T})^{-1/(q-1)}
\end{equation}
where $T$ is the effective temperature, $q$ is the entropy based parameter which characterizes the degree of deviation of the $p_T$ distribution from exponential Boltzmann Gibbs dsitribution. Eq. (2) is the non-extensive generalized form of the exponential Boltzmann distribution which introduces a new parameter $q$ to the temperature.

The Tsallis distribution function at mid-rapidity can also be expressed [57, 63, 64] as
  \begin{equation}
\label{eq:1}
\frac{1}{N}\frac{d^2N}{2\pi p_Tdp_Tdy} = Cm_T(1+(q-1)\frac{m_T}{T})^{-q/(q-1)}
\end{equation}
 which is a thermodynamically consistent Tsallis function.

 Eq. (1) and (2) are mathematically equivalent if $n=1/(q-1)$ and $p_0=nT_0$. {\bf It should be noted that here $T_0$ is the kinetic freeze-out temperature. $T$ used in Eq. (1), (2) and (3) is the effective temperature which can be expressed as $T$=$T_0$+$<\beta_T>$ [72]}. By inserting $p_0$ in eq. (1), we have
 \begin{equation}
\label{eq:1}
\frac{1}{N}\frac{d^2N}{2\pi p_Tdp_Tdy} = C(1+\frac{m_T}{nT_0})^{-n}
\end{equation}
According to ref. [67, 73, 74, 75], the transverse flow is introduced in Eq. (4)
\begin{equation}
\label{eq:2}
\frac{1}{N}\frac{d^2N}{2\pi p_Tdp_Tdy} = C(1+\frac{<\gamma_T> (m_T - p_T<\beta_T>)}{nT_0})^{-n}
\end{equation}
where $\gamma_t=1/\sqrt{1-<\beta_T>^2}$, $<\beta_T>$ denotes the average transverse flow velocity, $T_0$ represents the kinetic freeze-out temperature. In the present work, Eq. (5) is known as the Hagedron function with embedded flow [67, 73--76] and it covers the low as well as the high $p_T$ regions.
The Hagedron function with embedded flow in eq. (5) is modified as
\begin{equation}
\label{eq:3}
\frac{1}{N}\frac{d^2N}{dp_Tdy} = 2\pi p_T C(1+\frac{<\gamma_T> (m_T - p_T<\beta_T>)}{nT_0})^{-n}
\end{equation}
In the present work, we have used eq. (6) for the individual fit of the particles.
\\

{\section{Results and discussion}}
\begin{figure*}[tp]
\begin{center}
\hskip-0.153cm
\includegraphics[width=15cm]{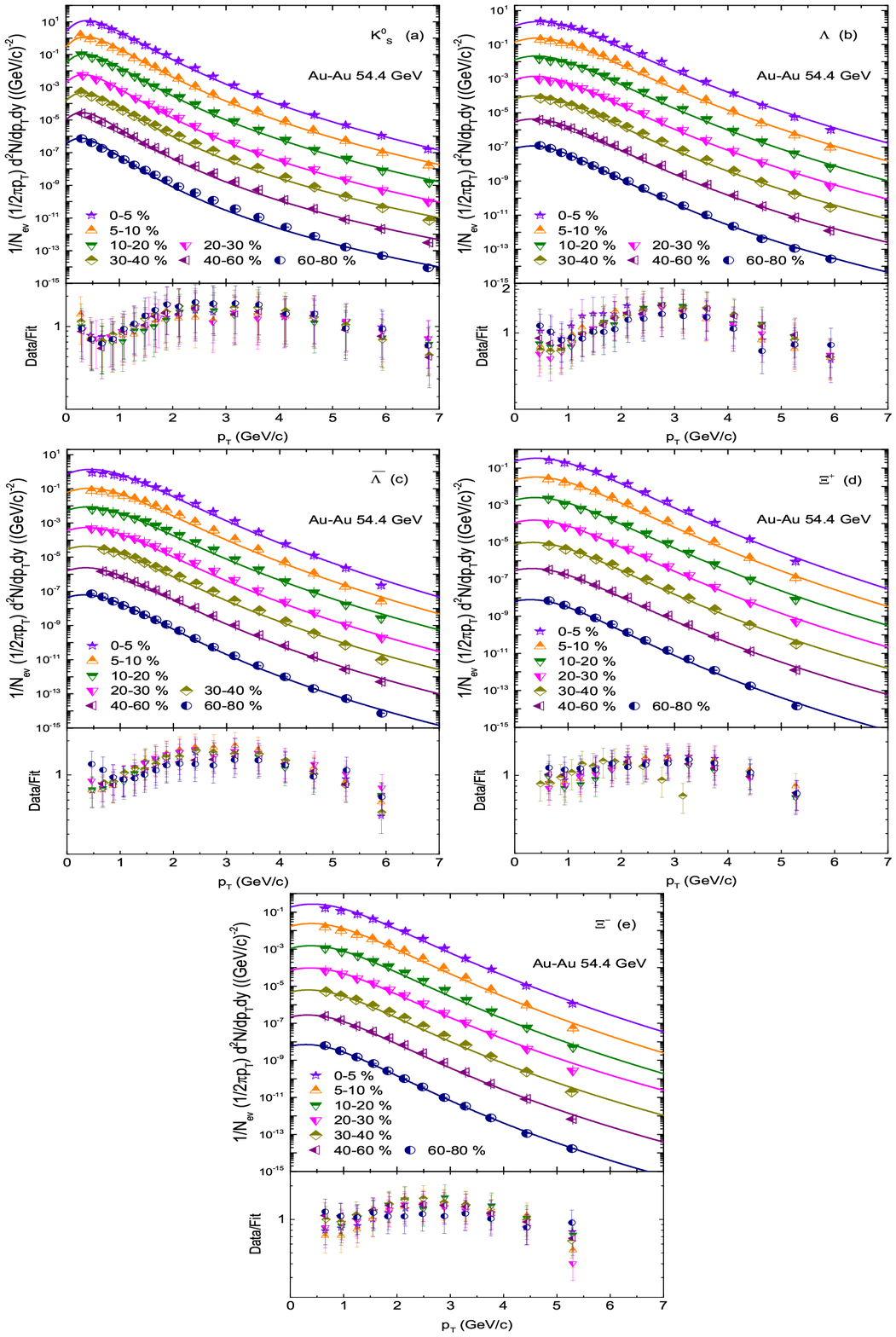}
\end{center}
Fig. 1. Transverse momentum spectra of $k_s^{0}$, $\Lambda$, $\bar {\Lambda}$, $\bar{\Xi}^{+}$ and $\Xi^{-}$ in Au-Au collisions at 54.4 GeV. The data points are the experimental data of STAR Collaboration [29] while curves show our fit results by eq. (6). Each panel consists its data/fit ratios in their lower part.
\end{figure*}

\begin{figure*}[ht!]
\begin{center}
\hskip-0.153cm
\includegraphics[width=15cm]{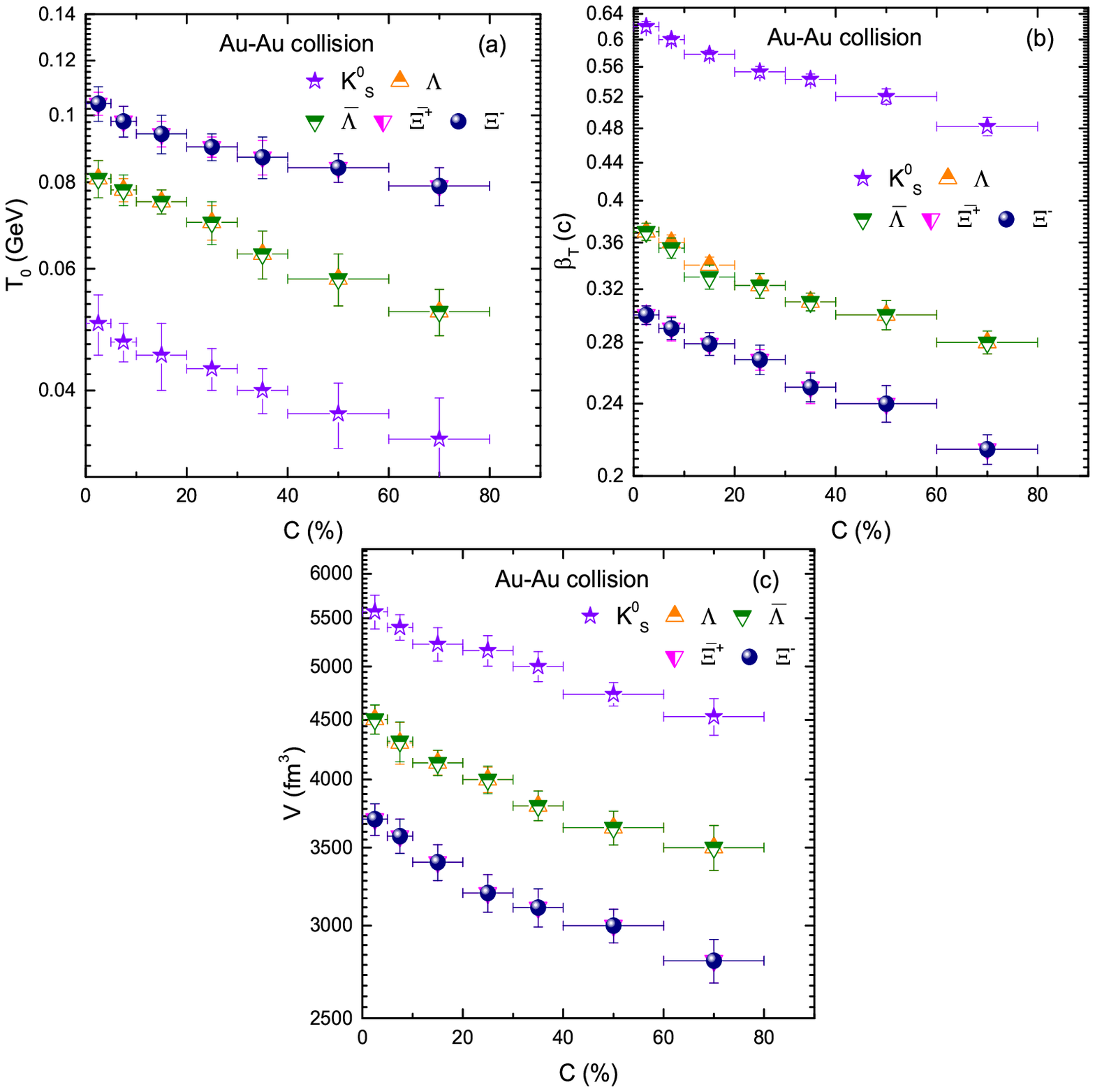}
\end{center}
Fig. 2. Dependence of (a) $T_0$, (b) $\beta_T$ and (c) $V$ on centrality as well as $m_0$.
\end{figure*}

\begin{table*}
\vspace{-1.50cm}
{\scriptsize Table 1. Values of free parameters $T_0$ and
$\beta_T$, V and n, normalization constant ($N_0$),
$\chi^2$, and degree of freedom (dof) corresponding to the curves
in Figs. 1. \vspace{-.50cm}
\begin{center}
\begin{tabular}{ccccccccccc}\\ \hline\hline
Collisions  & Centrality    & Particle   & $T_0$           & $\beta_T$        & $V (fm^3)$   &    $n$          & $N_0$         & $\chi^2$/ dof \\ \hline
   Fig. 1  & $0-5\%$       & $K^0_S$    &$0.050\pm0.005$  & $0.620\pm0.008$  & $5568\pm185$ & $15.2\pm0.5$    & $3\pm0.3$                 & 3/15\\
   Au-Au   & $5-10\%$      & $K^0_S$    &$0.047\pm0.003$  & $0.600\pm0.006$  & $5400\pm136$ & $14.0\pm1.6$    & $0.30\pm0.03$             & 5/16\\
   4.4 GeV & $10-20\%$     & $K^0_S$    &$0.045\pm0.005$  & $0.578\pm0.008$  & $5225\pm172$ & $13.3\pm1.6$    & $0.025\pm0.006$           & 6/16\\
            & $20-30\%$     & $K^0_S$    &$0.043\pm0.003$  & $0.553\pm0.008$  & $5158\pm155$ & $12.6\pm2$      & $0.0013\pm0.0003$         & 4/15\\
            & $30-40\%$     & $K^0_S$    &$0.040\pm0.003$  & $0.543\pm0.007$  & $5000\pm148$ & $11.8\pm1.4$    & $1E-4\pm3E-5$             & 9/16\\
            & $40-60\%$     & $K^0_S$    &$0.037\pm0.004$  & $0.520\pm0.010$  & $4733\pm109$ & $11.5\pm0.5$    & $6E-6\pm4E-7$             & 10/16\\
            & $60-80\%$     & $K^0_S$    &$0.034\pm0.005$  & $0.482\pm0.011$  & $4530\pm163$ & $11.0\pm1.2$    & $1.5E-7\pm5E-8$           & 8/16\\
    \hline
            & $0-5\%$       & $\Lambda$  &$0.081\pm0.005$  & $0.370\pm0.008$  & $4505\pm130$ & $18.5\pm2.1$    & $0.64\pm0.07$             & 12/14\\
            & $5-10\%$      & $\Lambda$  &$0.078\pm0.003$  & $0.360\pm0.007$  & $4300\pm176$ & $18.2\pm1.7$    & $0.07\pm0.004$            & 11/14\\
            & $10-20\%$     & $\Lambda$  &$0.075\pm0.003$  & $0.340\pm0.007$  & $4136\pm100$ & $17.7\pm1.2$    & $0.006\pm0.0005$          & 11.4/14\\
            & $20-30\%$     & $\Lambda$  &$0.070\pm0.004$  & $0.323\pm0.010$  & $4000\pm101$ & $16.3\pm1.4$    & $3.8E-4\pm4E-5$           & 13.5/14\\
            & $30-40\%$     & $\Lambda$  &$0.063\pm0.005$  & $0.310\pm0.007$  & $3800\pm111$ & $15.4\pm1.3$    & $2.7E-5\pm5E-6$           & 13/14\\
            & $40-60\%$     & $\Lambda$  &$0.058\pm0.005$  & $0.300\pm0.011$  & $3639\pm121$ & $14.8\pm1$      & $1.2E-6\pm3E-7$           & 9/14\\
            & $60-80\%$     & $\Lambda$  &$0.052\pm0.004$  & $0.280\pm0.008$  & $3500\pm154$ & $14.4\pm1.4$   & $3E-8\pm4E-9$             & 5/14\\
   \hline
            & $0-5\%$       & $\bar {\Lambda}$  &$0.081\pm0.005$  & $0.370\pm0.008$  & $4505\pm130$ & $20\pm2.3$    & $0.38\pm0.05$          & 30.7/14\\
            & $5-10\%$      & $\bar {\Lambda}$  &$0.078\pm0.004$  & $0.355\pm0.009$  & $4313\pm170$ & $18.6\pm1.5$  & $0.03\pm0.003$         & 23/14\\
            & $10-20\%$     & $\bar {\Lambda}$  &$0.075\pm0.003$  & $0.330\pm0.010$  & $4136\pm103$ & $17.0\pm1.2$  & $0.0024\pm0.0003$      & 16/14\\
            & $20-30\%$     & $\bar \Lambda$  &$0.070\pm0.005$  & $0.323\pm0.010$  & $4000\pm109$ & $16.5\pm1.5$  & $1.5E-4\pm3E-5$        & 123/14\\
            & $30-40\%$     & $\bar {\Lambda}$  &$0.063\pm0.005$  & $0.310\pm0.007$  & $3800\pm111$ & $15.4\pm1.3$  & $1.2E-5\pm4E-6$        & 25/14\\
            & $40-60\%$     & $\bar {\Lambda}$  &$0.058\pm0.005$  & $0.300\pm0.011$  & $3639\pm121$ & $15.2\pm1.1$  & $6.5E-7\pm4E-8$        & 10/14\\
            & $60-80\%$     & $\bar {\Lambda}$  &$0.052\pm0.004$  & $0.280\pm0.008$  & $3500\pm154$ & $15\pm1.2$    & $1.6E-8\pm5E-9$        & 8/14\\
   \hline
            & $0-5\%$      & $\bar \Xi^+$     &$0.104\pm0.004$  & $0.300\pm0.007$  & $3700\pm110$  & $22\pm2.8$   & $0.13\pm0.06$          & 10/9\\
            & $5-10\%$     & $\bar \Xi^+$     &$0.098\pm0.005$  & $0.290\pm0.009$  & $3580\pm125$  & $20.3\pm2.1$ & $0.0125\pm0.004$       & 6/9\\
            & $10-20\%$    & $\bar \Xi^+$     &$0.094\pm0.004$  & $0.279\pm0.008$  & $3400\pm120$  & $19\pm2.1$   & $0.001\pm0.0003$       & 7.5/9\\
            & $20-30\%$    & $\bar \Xi^+$     &$0.090\pm0.003$  & $0.268\pm0.007$  & $3200\pm140$  & $18\pm1.5$   & $6.2E-5\pm4E-6$        & 7/9\\
            & $30-40\%$    & $\bar \Xi^+$     &$0.087\pm0.005$  & $0.250\pm0.010$  & $3110\pm110$  & $17\pm1.1$   & $3.8E-6\pm3E-7$        & 7/9\\
            & $40-60\%$    & $\bar \Xi^+$     &$0.084\pm0.004$  & $0.240\pm0.011$  & $3000\pm100$  & $16.5\pm1.3$ & $1.5E-7\pm4E-8$        & 5/9\\
            & $60-80\%$    & $\bar \Xi^+$     &$0.079\pm0.005$  & $0.214\pm0.008$  & $2800\pm120$  & $16.4\pm1.3$ & $3E-9\pm3E-10$         & 6/9\\
   \hline
            & $0-5\%$      & $\Xi^-$     &$0.104\pm0.005$  & $0.300\pm0.007$  & $3700\pm115$  & $21.1\pm2.1$   & $0.11\pm0.04$             & 5.5/9\\
            & $5-10\%$     & $\Xi^-$     &$0.098\pm0.003$  & $0.290\pm0.008$  & $3580\pm120$  & $20.3\pm2.1$   & $0.0093\pm0.0006$         & 17/9\\
            & $10-20\%$    & $\Xi^-$     &$0.094\pm0.005$  & $0.279\pm0.008$  & $3400\pm120$  & $19\pm2.1$     & $6E-4\pm4E-5$             & 7.5/9\\
            & $20-30\%$    & $\Xi^-$     &$0.090\pm0.005$  & $0.268\pm0.010$  & $3200\pm118$  & $17\pm1.2$     & $4E-5\pm5E-6$             & 28/9\\
            & $30-40\%$    & $\Xi^-$     &$0.087\pm0.004$  & $0.250\pm0.009$  & $3110\pm117$  & $16.9\pm1.2$   & $2.5E-6\pm4E-7$           & 9/9\\
            & $40-60\%$    & $\Xi^-$     &$0.084\pm0.004$  & $0.240\pm0.011$  & $3000\pm100$  & $16.7\pm1.3$   & $1.1E-7\pm6E-8$           & 6.5/9\\
            & $60-80\%$    & $\Xi^-$     &$0.079\pm0.005$  & $0.214\pm0.008$  & $2800\pm120$  & $16.5\pm1.1$   & $2.7E-9\pm4E-10$          & 1/9\\
\hline
\end{tabular}%
\end{center}}
\end{table*}

\begin{figure*}[tp]
\begin{center}
\hskip-0.153cm
\includegraphics[width=15cm]{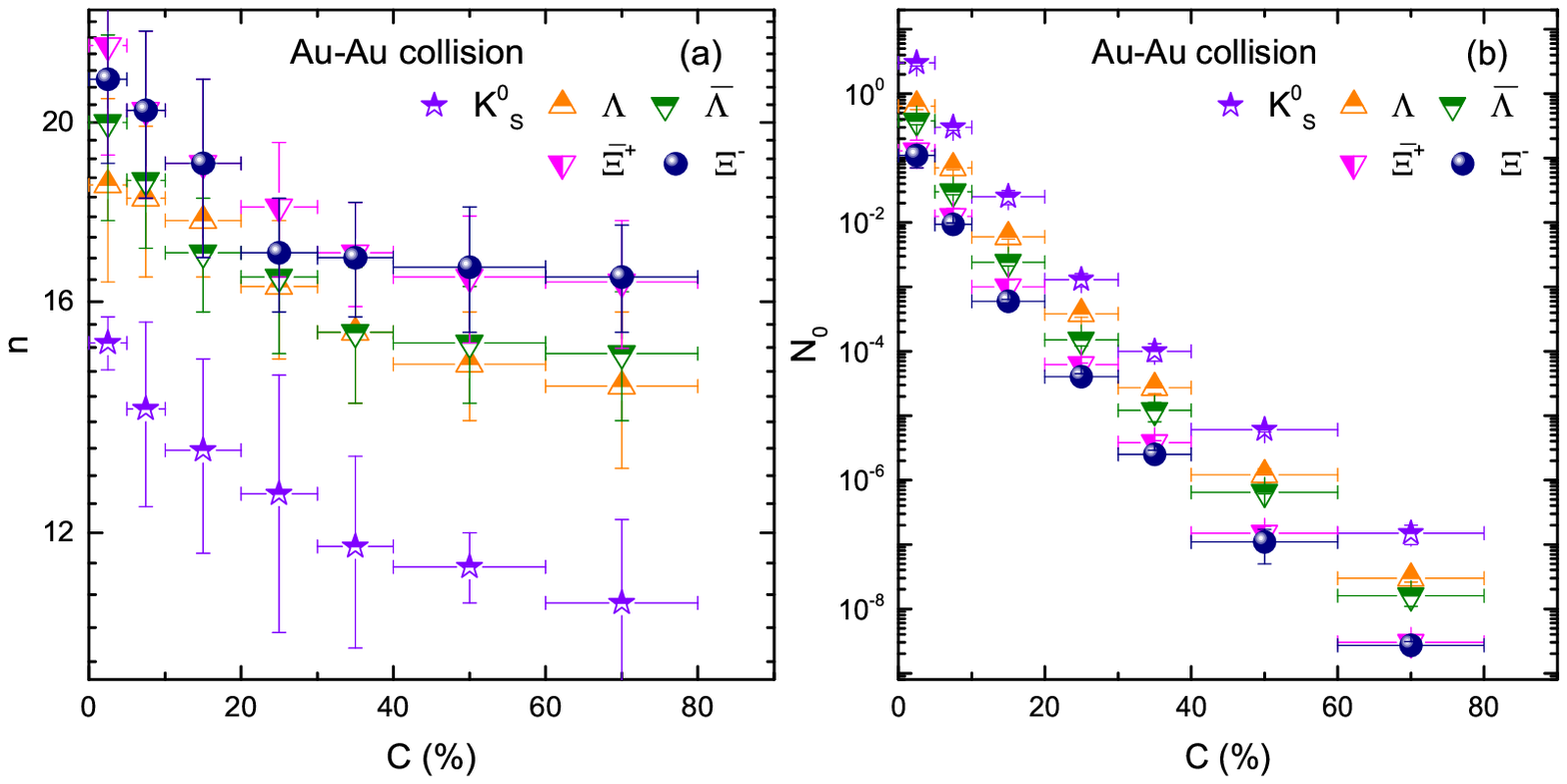}
\end{center}
Fig. 3. Dependence of (b) $n$, and (b) $N_0$ on centrality as well as $m_0$.
\end{figure*}

The STAR Collaboration [29] had reported the spectra of strange hadrons ($k_s^{0}$, $\Lambda$, $\bar {\Lambda}$, $\bar{\Xi}^{+}$ and $\Xi^{-}$) produced in Gold-Gold (Au-Au) collisions at $\sqrt {s_{NN}}$ = 54.4 GeV in various centrality intervals for mid-rapidity region of $|y| < 0.5$. Figure 1 panel (a)-(e) displays the $p_T$ spectra of $k_s^{0}$, $\Lambda$, $\bar {\Lambda}$, $\bar{\Xi}^{+}$ and $\Xi^{-}$ respectively in different centrality classes. The centrality classes include 0-5\%,5-10\%, 10-20\%, 20-30\%, 30-40\%, 40-60\%, 60-80\%. One can see different symbols in each panel which represents different centrality classes, while the curves over each data set in each panel are the results of our fitting by using the Modified Hagedorn function with embedded flow. The approximately well fit of the STAR data in mid-rapidity in Au-Au collisions by Eq. (6) can be seen. The lower panels show the data/fit ratios of the corresponding upper panel to monitor the quality of the fits. To confirm the firmness of the obtained parameter values, the fit procedure is repeated three times, changing the initial values of parameters. As a result, the stability of the obtained parameters (they have stayed practically the same) has been fully validated. The extracted values of the free parameters, normalization constant ($N_0$), $\chi^2$ and degrees of freedom (dof) are listed in table 1.

In order to explore more intuitively the dependence of the parameters on centrality, mass or their mutual correlation, we introduced fig. 2-7, where the parameters are cited from table 1.
Figure 2 shows the results of kinetic freeze-out temperature ($T_0$), transverse flow velocity ($\beta_T$) and kinetic freeze-out volume ($V$). In fig. 2 (a), $T_0$ as a function of centrality class and $m_0$ is displayed. Different symbols are used to represent different particle species, and the trend of the symbols from left to right show the trend of the parameters from central to peripheral collisions. It is reported that $T_0$ is larger for central collisions (0-5 \%) and decreases with decrease of centrality.

The central collisions correspond to the creation of higher concentration energies due to large number of participants involved in the reaction and hence it results in larger $T_0$. However, as the system moves toward periphery, the created concentration energies becomes less and less and thus it results in smaller $T_0$. This result is consistent with with our previous work [31, 50, 51, 54 , 76]. Ref. [31, 78] shows larger $T_0$ where $T_0$ is extracted by an alternative method in different collision systems for the non-strange particles. In Ref. [51] we analyzed non-strange, strange and multi-strange particles, and in ref. [50] light nuclei are studied by different models respectively in narrow $p_T$ range, and the result for $T_0$ is larger in central collisions. The increasing or decreasing trend of $T_0$ with centrality is a complex scenario. It sometimes depends on the model, sometimes it depend on other factors like the flow profile in some models and $p_T$ range. Besides, $T_0$ reveals the mass differential kinetic freeze-out scenario, as it is larger for the massive particles such that $k_s^{0}$ $<$ $\Lambda$ $<$ $\Xi$ (where $\Lambda$ stands for $\Lambda$, $\bar {\Lambda}$, and $\Xi$ stands for $\bar{\Xi}^{+}$ and $\Xi^{-}$). We also believe that this larger $T_0$ for $\Xi$ than $\Lambda$ and $k_s^{0}$, similarly larger $T_0$ for $\Lambda$ than $k_s^{0}$ maybe due to different quark contents or maybe quark flavors such as $d\bar s$ for $k_s^{0}$, $uds$ for $\Lambda$ and $ssd$ for $\Xi$. However to confirm it, we need more particles with different quark combinations and different quarks flavors, which we will do in future. It is noteworthy that $T_0$ for $\Lambda$, and $\bar {\Lambda}$, and for $\Xi$ and $\bar{\Xi}^{+}$ is equal which means that they have equal interaction with medium produced in the collisions respectively.

\begin{figure*}[tp]
\begin{center}
\hskip-0.153cm
\includegraphics[width=10cm]{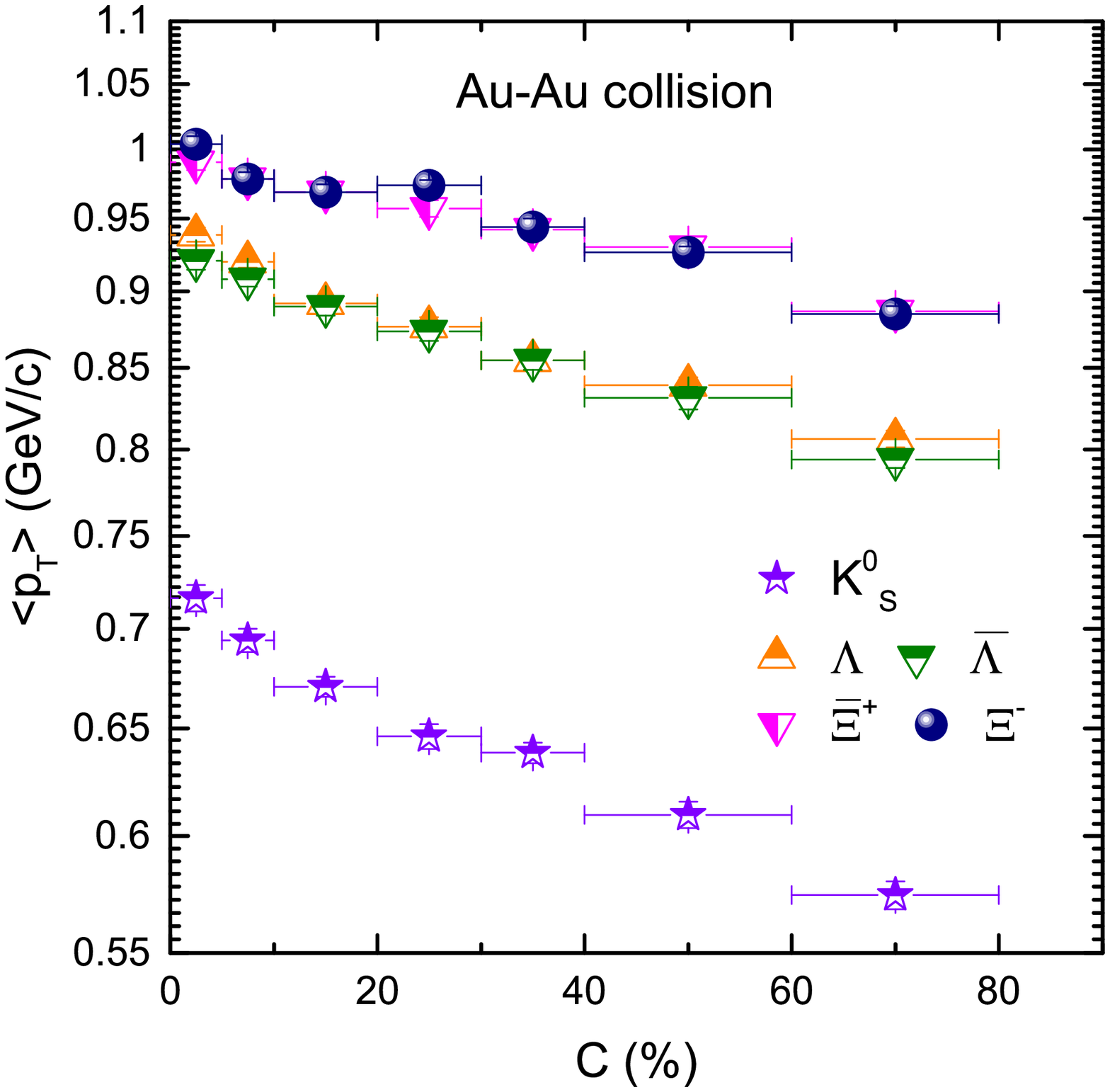}
\end{center}
Fig. 4. Dependence of $<p_T>$ on centrality as well as $m_0$.
\end{figure*}
Figure 2 (b) shows the transverse flow velocity $\beta_{T}$ dependence on centrality class and $m_0$. We reported that the collective behavior at the kinetic freeze-out stage
changes with centrality. The central collisions experience very harsh squeeze which deposits more energy per nucleon in the system that results in a rapid expansion of the fireball, while the squeeze becomes less violent as the system goes towards periphery and the amount of energy deposition in the system becomes less, and the expansion of the fireball becomes steady, which results in lower values of $\beta_T$ in peripheral collisions. We believe that the flow is produced in the inner core of the interacting system, therefore even for the peripheral collisions the flow velocity is non-zero. One can see that there exists a clear partition of $\beta_T$ among different centralities, such that $\beta_T$ is larger in most central collisions, while it is the lowest in most peripheral collisions, and it comes in between in intermediate centralities which renders that most central collision result in high pressure gradients and therefore larger velocities of produced particles flying out from this zone at expansion stage. We also reported that $\beta_T$ shows mass dependency. Lighter the mass of the particle, higher will be the transverse flow velocity and vice versa. The reason behind this is that the heavier particles are left behind in this system due to its large inertia and comes out later, however the light particles due to their small inertia come out of the system early.

Figure 2 (c) displays the kinetic freeze-out volume (V) as a function of centrality and $m_0$. Like $T_0$ and $\beta_T$, it also grows towards the central collisions. Large number of participants involved in interaction in central collisions which correspond to large number of binary
collisions by the re-scattering of partons that bring the system to equilibrium state quickly. While the number of participants reduces in
peripheral collisions and the system reaches to equilibrium state in a steady manner.
In addition to these observations, $V$ also depends on the mass of the particles. Lighter the mass of the particle, higher is the value of $V$, which renders the early freeze-out of massive particles. This indicates that there is a separate freeze-out surface for each particle and renders a volume differential freeze-out scenario. In the present work, $\Xi$ has the lowest $V$ which renders its early freeze-out.

\begin{figure*}[ht!]
\begin{center}
\hskip-0.153cm
\includegraphics[width=15cm]{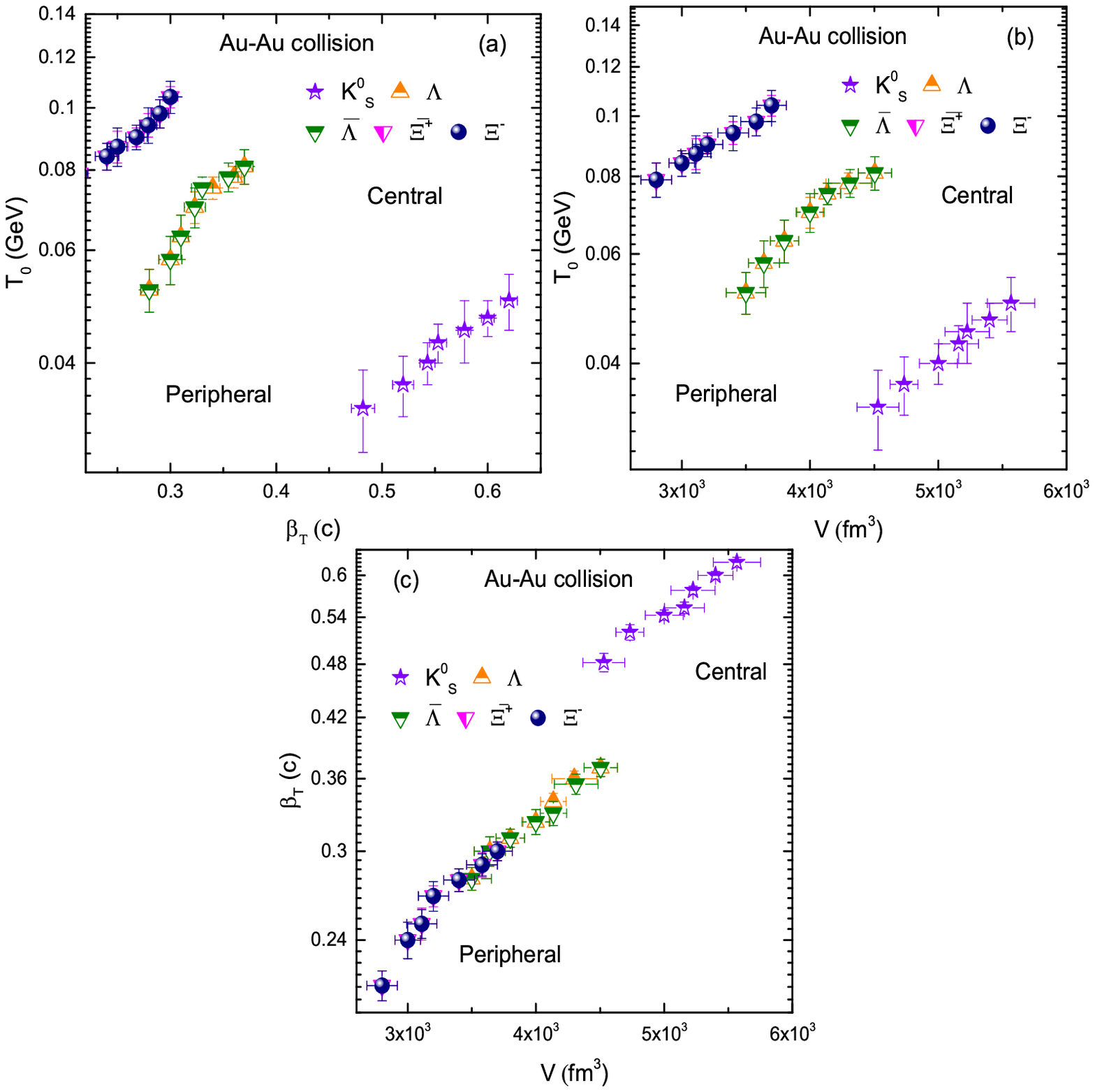}
\end{center}
%\vskip-2.0cm
Fig. 5. Correlation of $T_0$ versus $\beta_T$, $T_0$ versus $V$ and $\beta_T$ versus $V$.
\end{figure*}

Figure 3 shows the parameter $n$ and $N_0$ dependence on centrality and mass of the particle in Au-Au collisions for $k_s^{0}$, $\Lambda$, $\bar {\Lambda}$,$\bar{\Xi}^{+}$ and $\Xi^{-}$ strange particles. Fig. 3 (a) shows the result of the dependence of the parameter $n$ on centrality and $m_0$, where as fig. 3 (b) shows the result for the parameter $N_0$. In fig. 3 (a), the parameter $n$ is an entropy based parameter and is given as $n=1/(q-1)$. $n$ is referred for the degree of equilibrium or non-equilibrium. As a reciprocal of q, large $n$ corresponds to equilibrium. The more larger the value of $n$, more closer the system will be to equilibrium state. Usually, the parameter $q=1$ refers to equilibrium state, and $q$ $>$1 corresponds to non-equilibrium state. It can be seen that $n$ increases with the increase of centrality which means that the central collisions are more close to equilibrium state. Furthermore, $n$ is also observed to be dependent on the particle species, For instance, smaller values of $n$ can be seen for lighter particle ($K_s^{0}$), which indicates $K_s^{0}$ interaction with the medium produced in collisions is less and comes to equilibrium very steadily, however $\bar{\Xi}^{+}$ and $\Xi^{-}$ have larger values of $n$ which naturally means to interact more with the produced medium during the collision and therefore they have a quick approach to equilibrium state. It should be noted that the physical restrictions on the parameter values have been imposed by us during the fitting procedure. For instance $T_0$ for $K_s^{0}$ was allowed to vary between 0.034 to 0.050 GeV, and $n$ was restricted to the range from 11 to 15.2. Similarly $T_0$ for $\Lambda$ and $\Xi$ are restricted to change between 0.052 to 0.081 GeV, and 0.089 to 0.104 GeV respectively while $n$ changes from 14.4 to 20, and 16.4 to 22. 

\begin{figure*}[ht!]
\begin{center}
\hskip-0.53cm
\includegraphics[width=15cm]{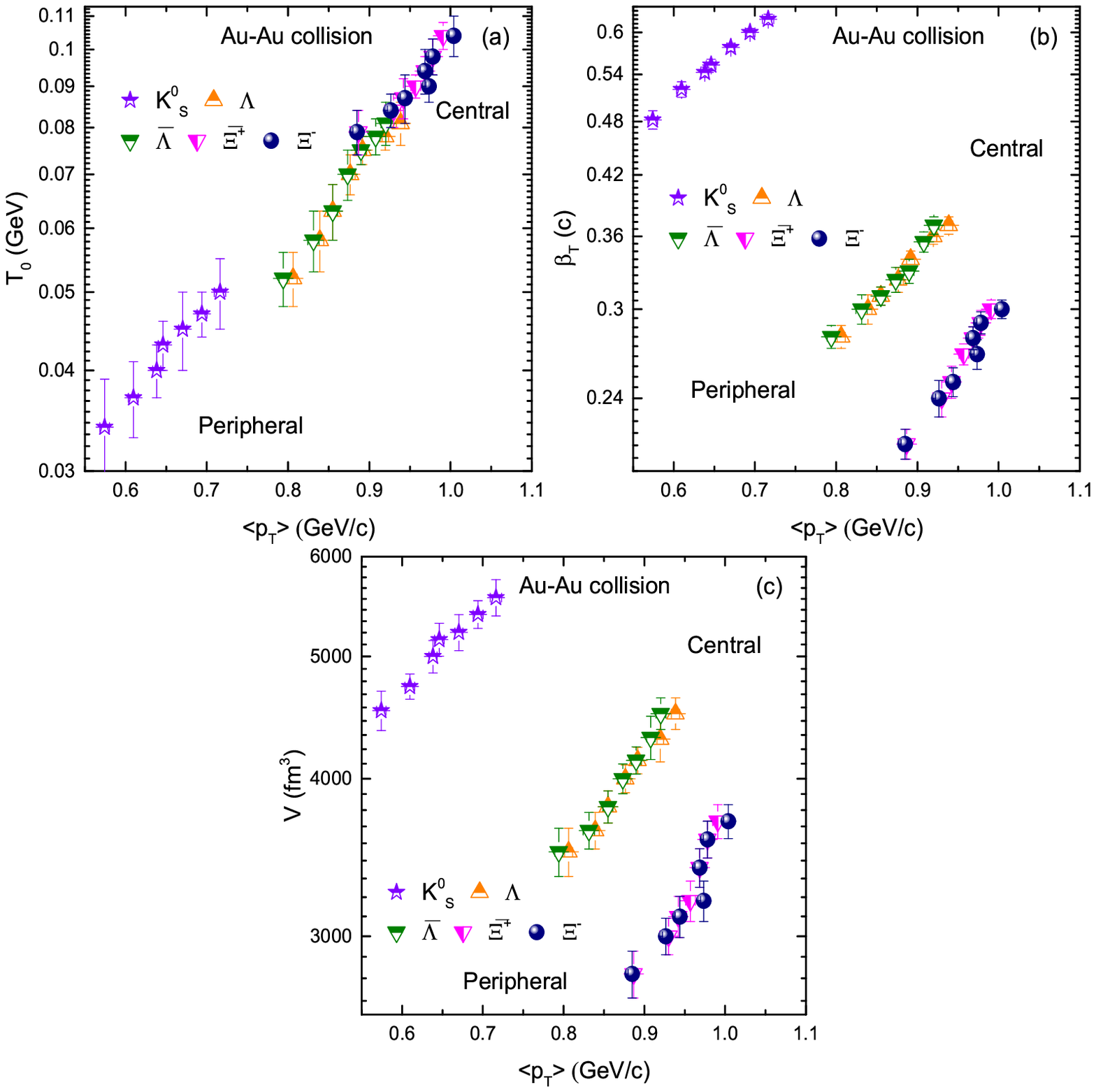}
\end{center}
%\vskip-2.0cm
Fig. 6. Correlation of (a) $T_0$ versus $<p_T>$, (b) $T_0$ versus $V$ versus $<p_T>$.
\end{figure*}
Fig. 3 (b) displays the dependence of the multiplicity parameter $N_0$ on centrality and $m_0$. {\bf $N_0$ is basically the normalization constant which is used to compare the fit function with the experimental spectra, but it has a physical significance. It shows the multiplicity. Here it should be noted that we discussed the normalization constant $C$ in Eq. (1), but $C$ is different from $N_0$, because $C$ is used to lead the integral of Eq. (1) to be normalized to unity. The sole purpose of the interpretation of both $C$ and $N_0$ is to give a clear description}. One can see an increasing trend of $N_0$ with increasing centrality. The fact is that in more central collisions more particles are involved in interaction which results in larger multiplicity after the collision reaction, where in peripheral collisions the participants in the interaction reduce and hence it results in lower rate of multiplicity. $N_0$ is also observed to be dependent on the mass of the particle. Massive particle corresponds to less $N_0$ which means that the multiplicity for massive particles is less. For instance in the present work, $\bar{\Xi}^{+}$ and $\Xi^{-}$ has lower multiplicity and $K_s^{0}$ has larger multiplicity, while $\Lambda$, $\bar {\Lambda}$ lies in between.

Fig. 4 is the same as fig. 2 and 3, but it demonstrates the dependence of mean $p_T$ ($<p_T>$) on centrality and $m_0$. We see that $<p_T>$ increases from $K^0_S$ to $\Lambda$ and then $\Xi$ which manifests that the  radial flow effect increases with particle mass. Besides, $<p_T>$ is observed to decrease with the decrease of event centrality which renders that the amount of energy gained by nucleons in the system in central collision is large that leads to further multiple scatterings. Whereas, the system in peripheral collisions has an opposite result. The larger $<p_T>$ in central collision is an indication of immense radial flow effects in more central collisions.
Fig. 5 shows the correlation of $T_0$ and $\beta_{T}$, $T_0$ and $V$ and $\beta_{T}$ and $V$.

\begin{figure*}[ht!]
\begin{center}
\hskip-0.153cm
\includegraphics[width=15cm]{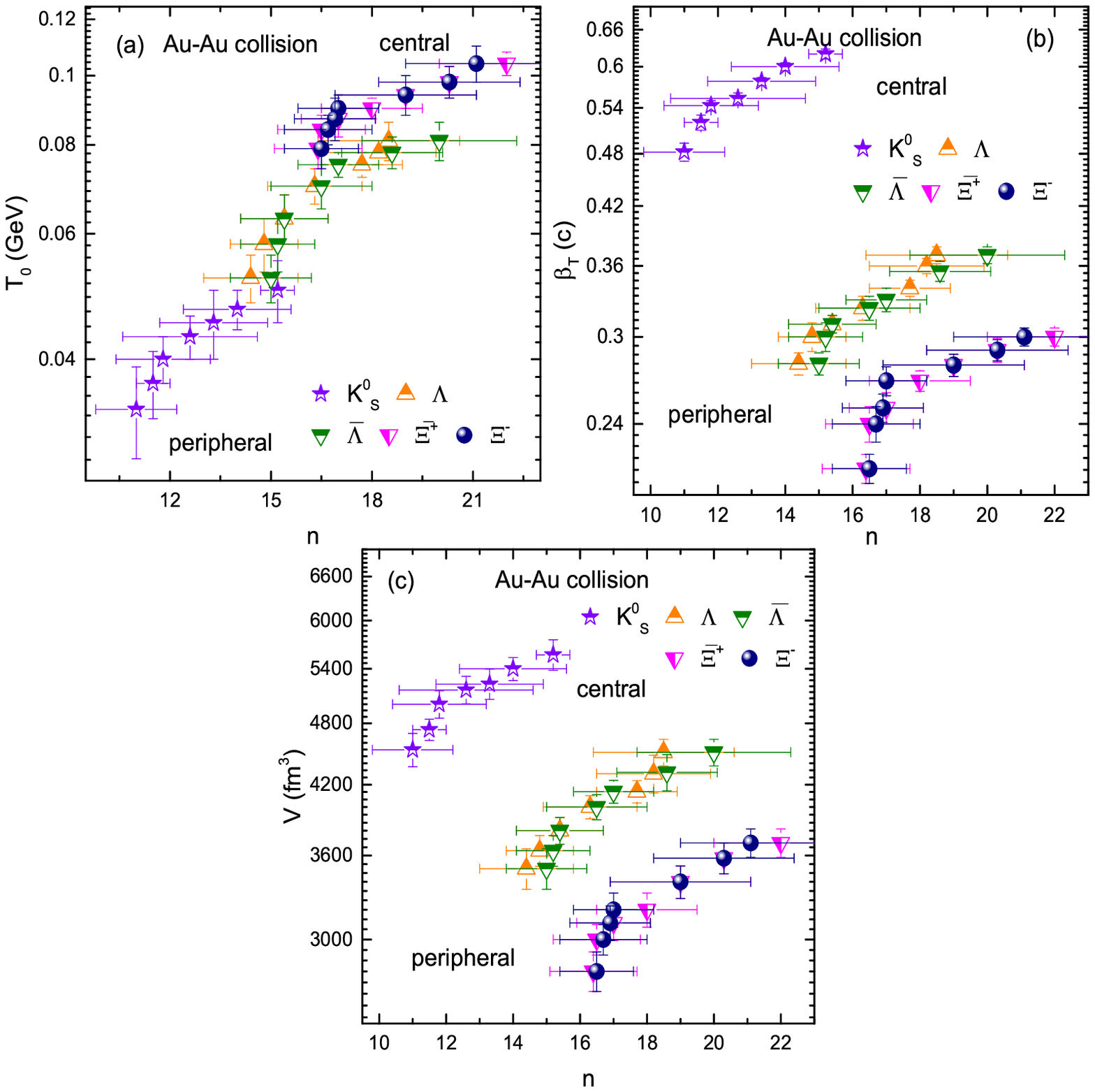}
\end{center}
Fig. 7. Correlation of (a) $T_0$ versus $n$, (b) $n$ versus $V$ versus $n$.
\end{figure*}

Fig. 5 (a) presents the correlation of $T_0$ and $\beta_{T}$ for all under studied particles. The symbols represent different particles and the trend of symbols from up to downward show the trend of correlation from central to peripheral collisions. One can see that there is a positive correlation between $T_0$ and $\beta_{T}$. $T_0$ increases with the increase of $\beta_T$. In central collision larger $T_0$ and $\beta_T$, which means that central collision systems gains higher degree of excitation which results in larger $T_0$ and a rapid expansion (harsh squeeze) which corresponds to larger $\beta_T$. This result validates our previous results [77]. Panel (b) shows the correlation of $T_0$ and $V$. It is observed that like $T_0$ and $\beta_T$, $T_0$ and $V$ also have a positive correlation. It can be explained as the higher degree of excitation results in larger $T_0$ due to violent collision as large number of participants are involved in interaction, which results in large multiplicity and hence lager $V$. While panel (c) shows the correlation of $\beta_T$ and $V$ and they also have a positive correlation. Because of very violent collisions, more energy is deposited in the system, which naturally results in larger multiplicity, and correspondingly the fireball expands rapidly (larger $\beta_T$).

Fig. 6 is similar to fig. 5, but it displays the correlation among $T_0$ and $<p_T>$, $\beta_T$ and $<p_T>$, and $V$ and $<p_T>$. Fig. 6 (a) shows the correlation of $T_0$ and $<p_T>$. One can see that the correlation between $T_0$ and $\beta_T$ is positive. Similarly in panel (b) and (c), the correlation between $\beta_T$ and $<p_T>$, and $V$ and $<p_T>$ is also positive. The positive correlation of $\beta_T$ and $<p_T>$ suggests that more energy is stored per nucleon in the system due to transfer of large amount of energy during collision which results in a rapid expansion of the system (large $\beta_T$). The parameters $T_0$ and $V$ are directly related to the freeze-out of the particles. Therefore the positive correlation of $T_0$ and $<p_T>$ renders that the larger $<p_T>$ (energy) transfer to the system occurs due to the violent collision that corresponds to a higher degree of excitation and results in larger $T_0$. We observed that in most central collisions of $\Lambda$, $\bar {\Lambda}$ coincides the most peripheral collisions of $\bar{\Xi}^{+}$ and $\Xi^{-}$ which may suggest the similar thermodynamic nature of central collision of $\Lambda$, $\bar {\Lambda}$ and peripheral collisions of $\bar{\Xi}^{+}$ and $\Xi^{-}$. On the other hand the positive correlation of $V$ and $<p_T>$ renders that larger energy transfer in the system results in large multiplicity which naturally results in larger $V$ if the density saturates. Or the second explanation maybe the larger transfer of energy results in longer evolution time and then larger partonic system, and hence larger $V$. The correlations in fig. 6 validate our above results.

Figure 7 is similar to fig. 6, but it presents the correlation of $T_0$, $\beta_T$ and $V$ with $n$. Panel (a) shows the correlation of $T_0$ and $n$,  panel (b) shows the correlation of $\beta_T$ and $n$, while panel (c) shows the correlation of $V$ and $n$. In all panels, a positive correlation among the relevant parameters can be seen.

The positive correlation of entropy based parameter $n$ with $T_0$ and $V$ renders that larger $T_0$ in central collisions can be explained as the short lived fireball in central collisions which results in the early freeze-out of the particles in central collisions, and the freeze-out happens when the system is in equilibrium which means larger $n$. In addition, central collisions are responsible for more number of particles to be involved in interaction where there are large number of binary scattering which results in large $V$ and large number of the system with large number of binary scatterings comes to equilibrium quickly and hence larger $n$. In the present we have seen that heavier particles has larger $T_0$, $n$ and smaller $V$ and hence they freeze-out earlier than the lighter particles. In addition, along with $T_0$ in central collisions $\beta_T$ is also larger and shows a rapid expansion of the fireball which means that the lifetime of the fireball is not long in central collisions compared to the peripheral collisions. Therefore in the present work, longer lived fireball corresponds to less $\beta_T$ and $T_0$ in peripheral collisions where the system is far from equilibrium and the particles freeze-out later.
\\
 {\section{Conclusions}}
 We summarize our main conclusions here

(a) We reported the transverse momentum spectra of $K_s^{0}$, $\Lambda$, $\bar {\Lambda}$, $\bar{\Xi}^{+}$ and $\Xi^{-}$ in different centrality intervals at 54.4 GeV. The modified hagedorn function with embedded flow is used to fit the experimental data of STAR Collaboration and obtained the freeze-out parameters by using the least square method.

(b) We obtained the results for kinetic freeze-out temperature ($T_0$), transverse flow velocity ($\beta_T$), kinetic freeze-out volume ($V$), entropy parameters ($n$), the multiplicity parameter ($N_0$) and mean transverse momentum ($<p_T>$). It is observed that all of these parameters are larger in central collisions and decrease towards periphery.

(c) We reported the decreasing trend of $T_0$ with the decrease of centrality due to the fact that the number of hadrons involved in interaction decrease from central to peripheral collisions, which correspond to a decrease in the degree of excitation of the system towards periphery and resultantly $T_0$ decreases.

(d) We also reported the decreasing trend of $\beta_T$ with the decrease of centrality due to the fact that central collision deposits more energy per nucleon in the system that results in rapid expansion of the fireball. Largest $\beta_T$ in central collisions show that there is a high pressure gradient in central collisions.

(e) A decreasing trend of $V$ with decreasing centrality is observed and the fact behind this is that the number of participants in the interaction decreases towards periphery and therefore the binary collisions by the re-scattering of partons decreases from central to peripheral collisions. $n$ is the smallest for $K_s^{0}$ which means that it interacts less with the medium produced in the collision.

(f) The entropy based parameter (n) is also studied and its decreasing trend from central to peripheral collisions is seen, which indicates the larger $n$ (smaller entropy) in central collision which is the evidence that central collisions are more close to the equilibrium state, while $n$ decrease towards periphery and the system goes away from equilibrium state.

(g) The multiplicity parameter ($N_0$) decreases from central to peripheral collisions because the central collisions are more harsh reactions and naturally they correspond to larger multiplicity, however it decreases towards periphery because the collisions becomes less violent towards periphery.

(h) We also calculated the mean transverse momentum which decreases from central collision to periphery due to the reason that in central collisions more energy is deposited in the system, while this deposition of energy decreases towards periphery. The decrease of $<p_T>$ towards periphery show that the radial flow effect decrease when the system moves toward the peripheral collisions. $<p_T>$ is also observed to be larger for the heavier particle which shows the effect of radial flow is larger in heavier particles. 

(i) $T_0$, $\beta_T$, $V$, $n$, $N_0$ and $<p_T>$ are mass dependent. $T_0$, $n$ and $<p_T>$ increase while $\beta_T$, $V$ and $N_0$ decrease with increasing the rest mass of the particle. The increase of $T_0$ and decrease of $V$ with $m_0$ indicate the multiple kinetic freeze-out and volume differential kinetic freeze-out scenarios respectively.

(j) We reported that there is a positive correlation among $T_0$ and $\beta_T$, $T_0$ and $V$, and $\beta_T$ and $V$. In addition, the correlation of $<p_T>$ with $T_0$, $\beta_T$ and $V$ is also checked and is seen to be positive. We also noticed the positive correlation of the entropy based parameter $n$ with $T_0$, $\beta_T$ and $V$.
In the correlation of $\beta_T$ and $V$, the most peripheral collisions of $\Lambda$, $\Lambda^{-}$ are seen to coincide the most central collisions of $\bar{\Xi}^{+}$ and $\Xi^{-}$. Similar is the case in the correlation among $T_0$ and $<p_T>$. This is maybe the indication of their similar thermodynamic nature.

   )
\\
\\

{\bf Data availability}

The data used to support the findings of this study are included
within the article and are cited at relevant places within the
text as references.
\\
\\
{\bf Compliance with Ethical Standards}

The authors declare that they are in compliance with ethical
standards regarding the content of this paper.
\\
\\
{\bf Acknowledgements}

The authors would like to thank support from the National Natural Science
Foundation of China (Grant Nos. 11875052, 11575190, and 11135011). Simultaneously, this work is partially supported by the Strategic Priority Research Program of Chinese Academy of Sciences (Grant No. XDPB15). We would also like to thanks the support from Ajman University International Research  Grant No. DGSR Ref. 2021-IRG-HBS-12, and Abdul Wali khan University, Mardan,
\\
\\

\end{multicols}
\end{document}